# Far-field Super-resolution Chemical Microscopy

*Mingwei Tang[1,2], Yubing Han[1], Danchen Jia[4], Qing Yang[1,3], and Ji-Xin Cheng[4],*

[1] State Key Laboratory of Modern Optical Instrumentation, College of Optical Science and Engineering, Zhejiang University, Hangzhou, 310027, China
[2] Research Center for Sensing Materials and Devices, Zhejiang Lab, Hangzhou, Zhejiang, 311121, China
[3] Research Center for Humanoid Sensing, Zhejiang Lab, Hangzhou, 311100, China
[4] Department of Biomedical Engineering, Department of Electrical and Computer Engineering, Photonics Center, Boston University, Boston, Massachusetts, 02459, United States

* Corresponding Author: Ji-Xin Cheng (jxcheng@bu.edu)

**Abstract:**

Far-field chemical microscopy providing molecular electronic or vibrational fingerprint information opens a new window for the study of three-dimensional biological, material, and chemical systems. Chemical microscopy provides a nondestructive way of chemical identification without exterior labels. However, the diffraction limit of optics hindered it from discovering more details under the resolution limit. Recent development of super-resolution techniques gives enlightenment to open this door behind far-field chemical microscopy. Here, we review recent advances that have pushed the boundary of far-field chemical microscopy in terms of spatial resolution. We further highlight applications in biomedical research, material characterization, environmental study, cultural heritage conservation, and integrated chip inspection.

## 1. Introduction

Breaking the optical diffraction limit has been a longstanding challenge to study the cellular- and molecular-scale activities of living systems and the nanoscale dynamics of novel materials. The 2014 Nobel Prize in Chemistry was awarded to Eric Betzig, Stefan W. Hell, and Willian E. Moerner in honor of their contribution to super-resolution fluorescence microscopy. Existing far-field super-resolution techniques have demonstrated remarkable success allowing for unprecedented spatial resolution below tens of nanometers, including stimulated emission depletion (STED) [1-3], single-molecule localization microscopy (SMLM) [4,5], and nonlinear structured illumination microscopy (Nonlinear-SIM) [6,7]. The working principles of these techniques, however, rely on the photo-physical properties of specially designed fluorophores. Although fluorescence microscopy brings the high contrast, signal-to-noise, and specificity that is useful for the investigator, it could not be applied for imaging objects that are neither autofluorescent nor fluorescence-tagged. In addition, the introduction of fluorescent labels onto the molecular structures of interest is likely to cause both functional and structural disruption of the target molecule, possibly leading to erroneous conclusions.

To circumvent the requirement of fluorescent labeling for long-duration imaging of living cells, label-free super-resolution microscopy using inherent physical properties from the samples instead of fluorescence markers as the contrast has been pursued, such as spatial frequency shift[8-11], microsphere lens[12-14], super oscillatory lenses[15, 16], and hyper lens[17-20]. Different from fluorescent super-resolution microscopy, label-free super-resolution microscopy is suitable for observing samples that are nonluminous and cannot be stained with fluorescent tags as in biological systems, such as carbon tubes, graphene, and integrated chips. However, these methods only provide morphological information, but not chemical selectivity. Super-resolution chemical microscopy is a useful tool that can provide valuable insights into the functionality and dynamic activities of cellular and molecular structures. By combining chemical selectivity and diffraction unlimited spatial resolution, this technique offers several advantages that can enhance our understanding of biological systems and functional materials.

Chemical microscopy has been extensively used in qualitative and quantitative analysis of biological specimens and novel materials by studying the interaction between light and matter without the requirement of fluorescent labeling. As shown in **Fig. 1**, the imaging contrast can come from the electronic-state or vibrational-state transition of the molecules, making chemical microscopy a non-invasive imaging technique with high molecular specificity. Vibrational-state chemical imaging senses either the infrared (IR) absorption or the Raman scattering [21], of which the representatives are Fourier-transform infrared absorption (FTIR) microscopy (together with its attenuated total reflection accessories), mid-IR photothermal (MIP) microscopy, mid-IR photoacoustic microscopy, spontaneous Raman microscopy, stimulated Raman scattering (SRS) microscopy, and coherent anti-Stokes Raman scattering (CARS) microscopy. Based on electronic transitions, transient absorption (TA) microscopy has been exploited to study ultrafast electronic-state dynamics of materials, such as graphene, using the mechanism of stimulated emission (SE), ground state depletion (GSD), or excited state absorption (ESA). Improving the spatial resolution is an essential step toward the ultimate goal of chemical microscopy of single molecule or single chemical bond. It is believed that studying the changes in chemical composition at the nano to meso length scales is vital for gaining insights into the structure, function, and interaction with the environment. Therefore, the significance of nanoscale chemical microscopy cannot be overstated as it is utilized in a variety of fields such as molecular biology, medicine, material science, and chemical science.

Compared with microscopy in the visible range, chemical microscopy, especially IR absorption-based CM, usually uses longer excitation wavelengths spanning from 0.5 μm to 10 μm, which makes it challenging to acquire high spatial resolution in chemical microscopy. The spatial resolution of an optical microscope can be defined by the Abbe diffraction limit of $\lambda/(2NA)$, where $\lambda$ is the wavelength of the light and NA is the numerical aperture of the focusing objective. Besides, the NA of the objective lens working in the IR spectrum is limited, making the spatial resolution typically to be larger than 3 μm (e.g., for $\lambda = 5$ μm and NA = 0.8). Near-field chemical imaging techniques are developed to solve this dilemma, by combining an atomic force microscope (AFM) with an infrared source[22-24]. In scattering-scanning near-field optical microscopy (s-SNOM), scattering from the AFM tip is collected[25]. In AFM-IR, absorption of IR radiation by the sample causes a thermal expansion, which deflects the AFM cantilever. The collection of high-spatial-frequency information by the near-field probe allows a spatial resolution of ~20 nm[26], which breaks the diffraction limit by ~150 times. Raman scattering-based near-field chemical microscopy was also developed. Tip-enhanced Raman scattering (TERS) is realized by combining tip-scanning imaging with Raman spectroscopy[27-30], where the metal tip not only provides a sub-10-nm high-resolution collection, but also enhances the Raman scattering signal by 4 to 7 orders. In these near-field modalities, however, the probe must be carefully controlled to be contacted with or 5-10 nanometers above the sample surface, which increases the complexity of the instrument and operation.

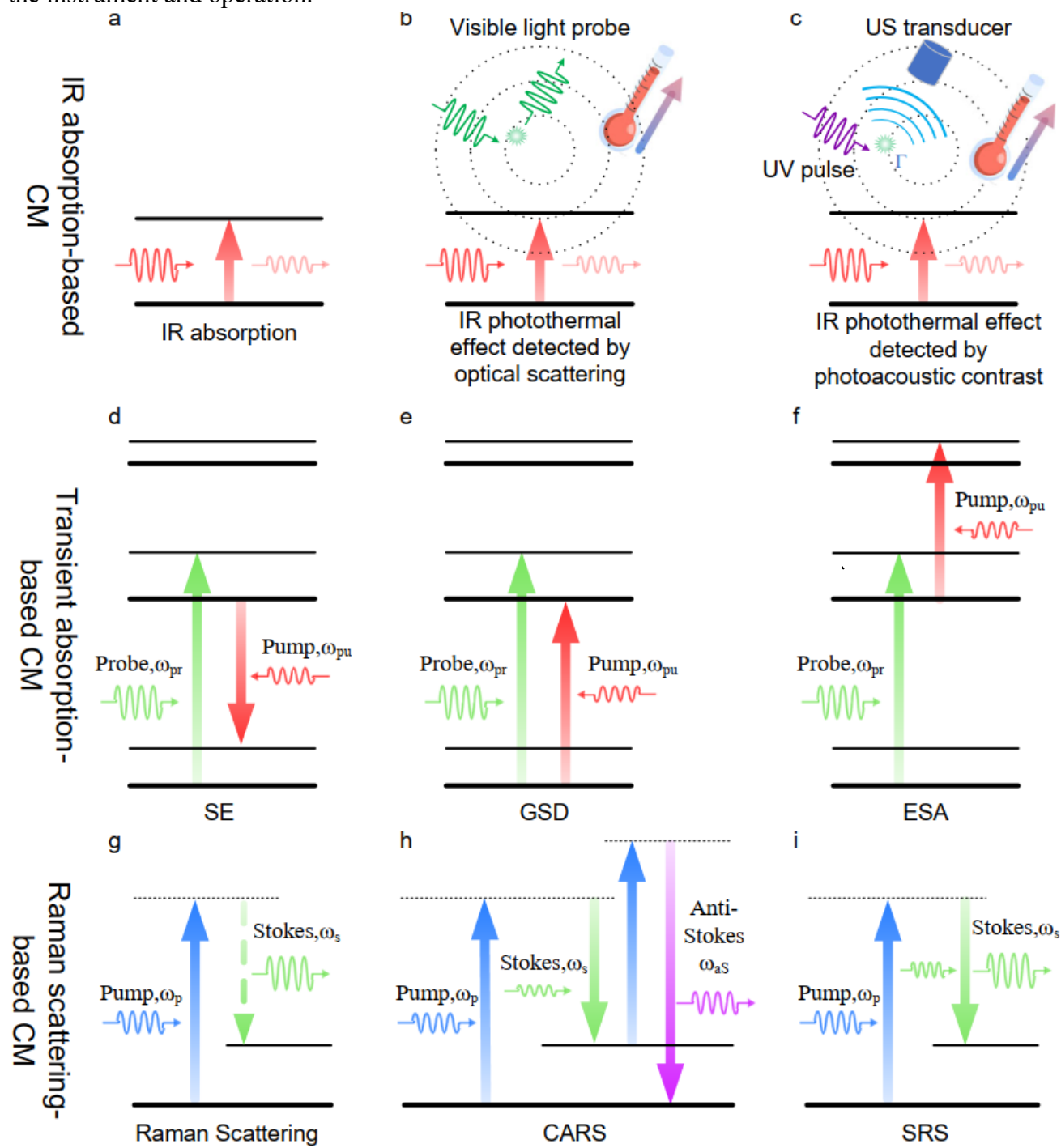

**Fig. 1 The principle and energy diagram of various chemical microscopies. a-c** IR absorption-based chemical microscopies. The principle of IR absorption (a), IR photothermal effect detected by optical scattering (b), and IR photothermal effect detected by photoacoustic contrast (c) is shown. **d-f** Transient absorption-based chemical microscopies. The principle of transient absorption based on stimulated emission (d), ground state depletion (e), and excited state absorption (f) is shown. **g-i** Raman scattering-based chemical microscopies. The principle of spontaneous Raman scattering (d), CARS (e), and SRS (f) is shown. CM: chemical microscopy.

Recently, the introduction of far-field super-resolution techniques endows chemical microscopy with resolution surpassing the diffraction limit, generating super-resolution chemical microscopy (SRCM). Owing to the chemical specificity and high sensitivity, SRCM can provide unique advantages in material science and biomedical applications, including nanomaterial inspection and living cell imaging. **Table 1** summarizes the SRCM methods discussed in this review.

**Table 1**. Summary of super-resolution chemical microscopies, including near-field and far-field techniques. SI: structured illumination. PA: photoacoustic. iSCAT: interferometric scattering. DH: digital holography. ODT: optical diffraction tomography. IDT: intensity diffraction tomography. DPP: antiphase demodulation pump-probe. SLI: structured-line illumination. SERS: surface enhanced Raman scattering. HO: high-order. ExM: expansion microscopy. LR: lateral resolution. AR: axial resolution.

| Method | Diffraction limited resolution | Resolution achieved | Imaging speed | Near-field or far-field | Applications | Ref. |
|---|---|---|---|---|---|---|
| AFM-IR or IR-sSNOM | 3-30 μm | 20-100 nm | - | Near-field | Stiffness map and IR absorption map of microbiology, cellular biology and nanophotonics | 24, 25 |
| Dark-field MIP | 5.1 μm | 610 nm (λ/9.2) | 134s/100 μm × 100 μm | 3D, Far-field | Living Cells; Drug distribution in cells | 31 |
| iSCAT-MIP | 4.4 μm | 510 nm (λ/11.3) | 0.8 ms/frame | 2D, Far-field | PMMA film; Living Cells | 32 |
| DH-MIP | 8 μm | 860 nm | 1s/100 μm ×100 μm | 2D, Far-field | Polystyrene and porous silica beads | 33 |
| ODT-MIP | 3.8 μm | LR: 380 nm AR: 2.3 μm | 12.5min/100 μm ×100 μm | 3D, Far-field | Cells | 34 |
| IDT-MIP | 4.6 μm | LR: 350 nm AR: 1.1 μm | ~20s/100 μm × 100 μm | 3D, Far-field | Cells | 35 |
| UV-localized MIP by PA sensing | 3.9 μm | 260 nm (λ/13.5) | ~1500s/100 μm ×100 μm | Far-field | Cells; Mouse brain slices | 36 |
| PEARL | 5.0 μm | 120 nm (λ/50.5) | - | Far-field | Cells | 37 |
| STED-TA | 345 nm | 250 nm (λ/3.3) | ~0.64s/100 μm ×100 μm | Far-field | Graphite; Nanoplatelets | 38 |
| STED-TA | 858 nm | 90 nm (λ/10) | - | Far-field | Single-Layer Graphene folding and defects | 39 |
| STED-TA | 453 nm | 36 nm (λ/25) | - | Far-field | Graphene nano-wrinkles | 40 |
| DPP-TA | 161 nm | 60 nm (λ/7.5) | 214s-3200s /100 μm ×100 μm | Far-field | Copper interconnections in CPU chips; Single-walled carbon nanotube; Boundary artifacts on monolayer graphene | 41 |
| Nonlinear differential TA | 380 nm | 185 nm (λ/3.7) | - | Far-field | CdSe nanobelts | 42 |
| SI-TA | 223 nm | 114 nm (λ/ | - | Far-field | Silicon nanowire | 43 |

| | | 5) | | | | |
|---|---|---|---|---|---|---|
| Near-field Raman | 450 nm | Sub-10 nm | - | Near-field | Carbon nanotubes | 27-30 |
| Visible CARS/SRS | 240 nm | 130 nm (λ/3.5) | 140s-640s /100 μm ×100 μm | Far-field | Polymer beads; Neurons; Cells, Brain tissue section | 44, 45 |
| SLI-Raman | 495 nm | 275 nm (λ/2.1) | 100min/28.6 μm ×44.3 μm | Far-field | Polymer beads; Carbon materials; Brain slice | 46 |
| SI-Raman | 190 nm | 80 nm (λ/6.7) | >9s/32 μm ×32 μm | Far-field | Carbon nanotubes; Graphene | 47 |
| SI-SERS | 370 nm | 89 nm (λ/5.3) | - | Near-field illumination and far-field detection | Cells | 48, 49 |
| Saturated-SRS | 377 nm | 255 nm (λ/3.6) | ~2.5s/100 μm× 100 μm | Far-field | Living cells | 50 |
| HO-CARS | 328 nm | 196 nm (λ/2.8) | ~1s/100 μm×100 μm | Far-field | Cells | 51 |
| SMLM-SERS | 347 nm | 10-30 nm (λ/17.7 − λ/66) | ~150s/8 μm×6.4 μm | Near-field illumination and far-field detection | Collagen fibers; Cells | 52-54 |
| Depletion-Raman | 1.37 μm | 0.93 μm (λ/0.8) | - | Far-field | Diamond | 55 |
| ExM-SRS | 382 nm | 78 nm (λ/10) | - | Far-field | Cells | 56 |

Of these modalities, Raman scattering-based far-field SRCM is developed to break the resolution of Raman microscopy. Super-resolution methods based on different mechanisms have been applied to Raman microscopy, such as SIM, SMLM, STED, nonlinearity, and ExM. These methods lead to the development of SLI-Raman [46], SI-Raman[47], SI-SERS[48, 49], SMLM-SERS[52-54], STED-SRS[55], Nonlinear SI-CARS[57], Saturated-SRS[50], HO-CARS[51], and ExM-SRS[56].

Compared with Raman scattering, IR absorption possesses a larger cross-section and is highly sensitive in the fingerprint region. A pump-probe technique termed as MIP[31-34, 36] (or optical photothermal IR) microscopy was developed to break the infrared diffraction limit, where the energy absorbed by the mid-IR excited molecules is nonradiatively transformed into heat, which changes the local refractive index. Then, the optical-path-length change is measured with a visible probe light.

For electronic-state transient absorption microscopy, the depletion effect, nonlinearity, and structured illumination are applied to realize a much higher resolution. These methods lead to the development of STED-TA[38-41], Nonlinear differential TA[42], and SI-TA[43].

In this review, we summarize recent advances in far-field super-resolution chemical microscopy and discuss the potentials in further pushing the spatial resolution of chemical microscopy. In the following sections, we first discuss the principles and optical implementation of various far-field SRCMs. Next, we summarize the applications of far-field SRCMs in biomedical study, material study, and integrated chip inspection. Finally, we discuss potentially progressive approaches and remaining challenges in this ever-growing and fast-developing field.

## 2. IR-absorption super-resolution chemical microscopy

Mid-infrared (MIR) spectroscopic imaging is commonly performed by using an FTIR spectrometer equipped with an infrared focal-plane array detector[58]. However, the spatial resolution of FTIR imaging is constrained by the diffraction limit of IR photons. In addition, strong water absorption restricts the application for observing subcellular structures in living biological samples. Although the resolution can be improved by using solid immersion lenses[59, 60], the resolution can only reach around λ/2.6, which is not enough to study the subcellular structures and activities in living systems. The recent development in photothermal IR microscopy fills this gap.

Photothermal spectroscopy was reported in the 1970s to detect the thermal lensing effect induced by the absorption of the pump beam at focus using a probe beam[61]. Photothermal microscopy showed superb detection sensitivity that allows single nanoparticle (1.4 nm diameter) [62, 63] and single molecule imaging[64]. Photothermal microscope in the visible spectrum has achieved a resolution of 90 nm by probing the nonlinear photothermal lens effect[65]. The mid-infrared photothermal (MIP) microscope recently developed by the Cheng group[31, 32] and the Hartland group[66] breaks the infrared-wavelength diffraction limit with a pump-probe configuration, that is, using shorter-wavelength probe light to detect the temperature rise induced by IR absorption.

Unlike the direct measurement of the absorption in IR spectroscopy, a MIP microscope detects the photothermal effect caused by a local temperature rise ($\Delta T$) in the sample, which leads to local refractive index change $\Delta n$ and thermal expansion $\Delta \ell$:

$$\Delta n = \frac{dn}{dT}\Delta T$$

$$\Delta \ell = \frac{1}{\ell}\frac{d\ell}{dT}\ell \Delta T$$

The scattered field $E_s$ from the sample under the probe field $E_i$ can be expressed as $E_s = |s(n,l)|e^{i\varphi_s(n,l)}E_i$, where $|s(n,l)|$ and $\varphi_s(n,l)$ are the amplitude and phase of the scattered field. With the modification of $\Delta n$ and $\Delta \ell$, the scattered field experiences changes in the intensity and phase delay, which could be detected through measurement of dark-field scattering [31, 67], optical phase[34, 35], and interferometric scattering[68, 69]. Below we discuss the various MIP configurations (**Fig. 2**) and the optical resolutions achieved by these methods.

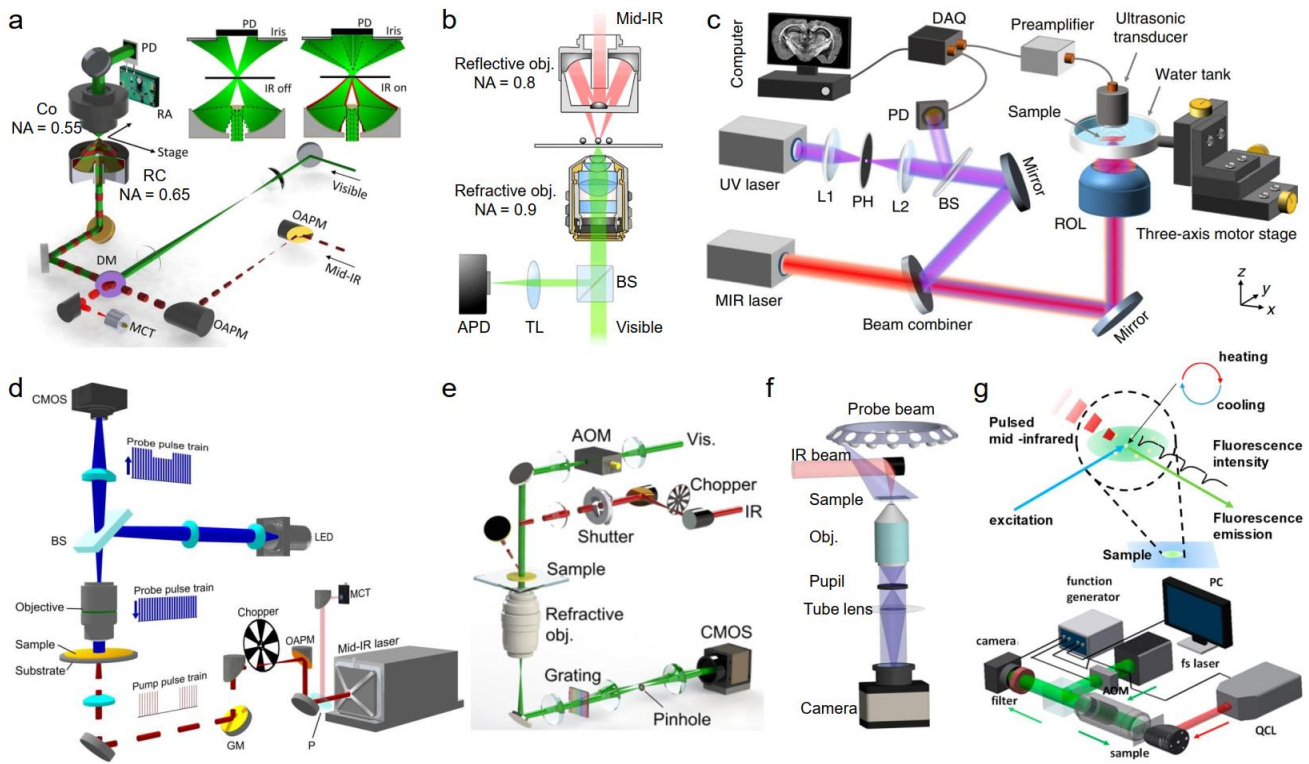

**Fig. 2 Implementation of various MIP microscopies. a-c** Scanning confocal MIP systems. (a)Co-propagated confocal MIP microscopy. Reproduced with permission[31]. Copyright 2016, the Authors, published by AAAS. (b)Counter-propagated confocal MIP microscopy. Reproduced with permission[66]. Copyright 2017 American Chemical Society. (c)UV-localized MIP photoacoustic microscopy. Reproduced with permission[36]. Copyright 2019 Nature Publishing Group. **d-g** Widefield MIP systems. (d)Widefield reflected interferometric scattering measurement. Reproduced with permission[32]. Copyright 2019, the Authors, published by AAAS. (e)Widefield phase measurement. Reproduced with permission[34]. Copyright 2020 Optical Society of America. (f) 3D phase measurement. Reproduced with permission[35]. Copyright 2022, the Authors, published by Nature Publishing Group. (g) Fluorescence-detected MIP measurement. Reproduced with permission[70]. Copyright 2021 American Chemical Society.

## Point scanning MIP microscope

*Visible-probed MIP.* With the deployment of a visible probe beam at 785 nm, Zhang and colleagues achieved sub-micrometer (~0.6 μm) chemical imaging of living cells or microorganisms via a confocal mid-IR photothermal microscope with a Cassegrain objective (0.65 NA)[31]. As shown in **Fig. 2**a, the visible probe beam and mid-IR pump beam are collinearly combined by a silicon dichroic mirror and directed to an inverted microscope. The visible signal light is collected in the transmission mode by another condenser. To increase the imaging SNR, a resonant amplifier that selectively probes the photothermal signal at the repetition rate of the pulsed IR laser is exploited. Since the probe focus is ~1/10 of the size of the pump mid-IR beam, objects outside the probe beam focus are not detected, leading to a nine-fold improvement of the spatial resolution. The confocal configuration also brings the advantages of optical sectioning capabilities. 3D imaging of living cells using infrared spectroscopy was demonstrated. For opaque samples, the visible signal light could also be detected in the epi-mode by the same Cassegrain objective. With this backward-detected MIP system[71], chemical mapping of active pharmaceutical ingredients and excipients of drug tablets has been demonstrated.

An alternative approach is to deploy a high NA objective for the counter-propagating probe beam. As shown in **Fig. 2**b, Li et al. realized an ultimate resolution of 300 nm using a 532 nm probe beam and a 0.9 NA objective in their counter-propagating MIP microscope[66]. The main advantage of this approach is that the visible probe beam can go through a high-NA refractive objective instead of low-NA reflective objectives that are typically used for IR sources, which improves the lateral resolution to ~300 nm. The forward-detection mode in the counter-propagation configuration is more sensitive to transparent samples, including cells and tissues, while the epi-detection mode in the co-propagation configuration is more suitable for small samples, such as bacteria and subcellular structures.

*UV-localized photoacoustic-sensing MIP.* The spatial resolution of MIP imaging can be further improved by UV photoacoustic detection[36]. Besides the refractive index change in photothermal effect, the Grüneisen parameter ($\Gamma$) of the material also changes with respect to the local temperature, leading to IR modulation of photoacoustic signals. As shown in **Fig. 2**c, the IR and UV light co-propagate through the samples immersed in the water tank, and the photoacoustic signals are detected by an ultrasound transducer. Since the photoacoustic signals are excited by UV light, a high lateral spatial resolution of 260 nm is achieved. Notably, this method lacks optical sectioning capability.

*Photothermal relaxation localization (PEARL) microscopy.* The resolution of the optical diffraction limit could be further broken by temporally probing the spatially inhomogeneous photothermal relaxation, which is recently presented by Fu et al. [37]. This method was a further achievement in pushing the resolution of MIR photothermal microscopy and it does not require special absorbers as it relies on general absorption processes such as electronic (E) and vibrational (V) absorption.

The photothermal process stimulated by the MIR pulse involves a heating phase due to vibrational absorption and an energy dissipation phase due to heat dissipation and mechanical waves. Since more energy was converted into acoustic waves towards the edge due to expansion, the resulting temperature increase is different in space and time even for homogenous absorbers. Therefore, a higher frequency is probed at the edge when sensed by the probe beam focus, showing a faster depletion. The difference in frequency of the center and edge could be detected with a conventional lock-in amplifier. The high resolution is closely related to the higher harmonic orders since the center-to-edge ratio shows an increasing trend at higher harmonic orders. By extracting high-harmonic components from the fast Fourier transform (FFT) of the temperature profile, the resulting images show a substantial improvement in resolution. In this work, the label-free bond-selective E- PEARL imaging of Gold nanoparticles with a resolution of 280 nm and V-PEARL imaging of living cells with a resolution of 120 nm (at the tenth harmonic) have been successfully demonstrated.

**Widefield MIP microscope**

With the scanning MIP systems, there remain a few restrictions. First, most IR photons do not contribute to the signal because of the mismatch of the IR and visible focal spot size. Second, the slow speed of sample scanning (typically ~20 s for 200×200 pixels[31]) makes it challenging to capture moving objects or for high-throughput detection. Below we discuss recent advances in improving the imaging speed and resolution by transforming point scanning into widefield schemes [32].

*Reflected interferometric scattering MIP.* The first widefield MIP microscope adopts a counter-propagation beam configuration[32]. As indicated in **Fig. 2**d, a 4-f lens system is used to project a 450-nm LED emitter to the objective back focal plane to create a uniform sample illumination in epi-configuration. The modulated IR pump beam is weakly focused on the sample plane with a CaF$_2$ lens, and the illuminated area is around 40 μm in diameter. The sample is prepared on a silicon wafer substrate to reflect the forward-scattered visible photons, providing a reference field for interferometric scattering measurement and good thermal conductivity for fast imaging. The reflected and scattered visible photons are collected with the same 0.66-NA objective and then projected on the camera with a tube lens. A spatial resolution of 0.51 um has been obtained thanks to the short wavelength probe light and high NA objective lens. Considering the multiplex advantage in wide-field MIP, for a 200×200 pixels image, the imaging speed is about 25 times faster than the point-scanning method.

*Phase-contrast MIP.* Since the MIP process involves the change of sample refractive index and dimension, which provide optical phase contrast, widefield phase imaging with chemical bond information has been demonstrated. By simply illuminating the sample with an intensity-modulated MIR wavelength tunable laser, specific molecular distribution can be obtained by subtracting phase maps obtained by a phase contrast microscopy in the cases of MIR-on and MIR-off. These phase contrast imaging methods, however, only provide qualitative phase distribution information and have inherent disadvantages such as limited resolution, shade-off effect, and contrast inversion, which will bring additional errors. Quantitative phase imaging (QPI) [72], such as digital holography (DH) [73, 74], optical diffraction tomography (ODT) [75, 76], intensity diffraction tomography (IDT) [77-79], yields sample-specific 2D optical-phase delay or 3D refractive-index distribution, which are the fundamental quantities used to visualize the morphology of transparent samples as in the cases of phase contrast methods. The combination of QPI and MIP provides complementary information about the sample: the chemical specificity and quantity of each molecular constituent. Recently, DH-MIP[33, 34](**Fig. 2**e), has been demonstrated to show these advantages in capturing the 2D distribution of protein or lipids of samples from living cells to fixed microorganisms. The axial and lateral bandwidths could be expanded simultaneously into a 3D synthetic aperture, resulting in depth- and super-resolved imaging performance. The 3D phase information could be extracted from scattered light fields of the sample interacted with the angle-varying oblique illuminations. The scattered light fields could be acquired either in the interferometrical or non-interferometric way, which leads to different methods, ODT-MIP[34] or IDT-MIP[35](**Fig. 2**f). The spatial resolution of these methods is mainly determined by the illumination NA and detection NA of the phase contrast realization. Up to now, ODT-MIP[34] can acquire a spatial resolution of 380 nm in the lateral and 2.3 μm in the axial. The imaging speed (12.5 mins per volume) however, is limited by the long averaging time to achieve an adequate signal-to-noise ratio, since the phase noise, optical misalignment, and mechanical instabilities will reduce the detection sensitivity. The IDT-MIP[35] method, recently presented by Cheng's group, performs with a higher throughput of ~20 s per volume, with a lateral and axial resolution of ~350 nm and ~1.1 μm, respectively.

*Fluorescence-detected MIP.* Although with great success, the sensitivity of label-free MIP imaging suffers from the weak dependence of scattering on the temperature, and the image quality is vulnerable to the speckles caused by scattering. To alleviate these shortcomings, MIP could be combined with thermosensitive fluorescent probes for the detection of photothermal effect with a much higher sensitivity[70]. As indicated in **Fig. 2**g, an infrared pulse train heats the surrounding of the fluorescent probe and causes a temperature rise, which subsequently modulates the fluorescence emission efficiency, then the fluorescence light instead of scattering light can be detected using a lock-in amplifier. The temperature-dependent emission efficiencies of common fluorophores including FITC, Cy2, Cy3, Rhodamine, and green fluorescent proteins are nearly 100 times compared with temperature-dependent scattering such that the imaging speed or SNR can be boosted by nearly 2 orders of magnitude. Taking advantage of the fluorescence signals, this method is compatible with super-resolution techniques used in fluorescence microscopies, such as SIM, STED, and STORM, providing future potentials in super-resolution MIP imaging through the thermal-sensitive fluorescent dye.

Up to now, the highest spatial resolution that has been achieved in widefield mid-IR photothermal microscopy is 0.35 μm, using 450 nm light illumination and an objective with 0.65 NA[35]. Further pushing the resolution limit could be realized by combining MIP with label-free super-resolution methods, such as spatial frequency shift[8, 9, 80-85]. With the illumination of a large lateral wavevector generated by high-refractive-index chips[10, 11], the resolution of linear label-free microscopy can be improved to sub-100 nm. Applying these methods to MIP microscopy could push the spatial resolution of IR imaging to break the visible-wavelength diffraction limit.

## 3. Transient absorption super-resolution chemical microscopy

Pump-probe chemical microscopy is a powerful tool for studying nonequilibrium dynamics in a variety of complex materials systems including nanostructures, low-dimensional semiconductors, and material interfaces. The simultaneous sub-micrometer spatial resolution and femtosecond temporal resolution enabled by the technique provides detailed, structurally correlated insights into intraparticle heterogeneity, free carrier transport, surface plasmon propagation, and unique interfacial states, which are inaccessible with conventional spectroscopies or microscopies.

Label-free absorption spectroscopies are frontline techniques to reveal the spectral fingerprint, composition, and environment of materials and are applicable to a wide range of samples. In an effort to improve the spatial resolution of far-field absorption microscopy, which is limited by the diffraction of light, an imaging technique based on transient absorption saturation has recently been developed.[38] Transient absorption microscopy is implemented by a pump-probe system with contrast mechanisms including stimulated emission, ground-state bleaching, and excited-state absorption.

**Depletion-based TA microscopy**

Depletion-based TA microscopy[38] is designed to decrease the probe area to below the diffraction limit in a pump-probe microscope by collinearly adding a non-modulated saturation beam, which has the same wavelength as the pump beam but with a much higher intensity. As shown in **Fig. 3**a, in the doughnut-shaped region of the focus where the intensity of the saturation beam is high, the transmission of the probe beam remains unchanged due to the saturation of the electronic transition. Under such conditions, the pump-to-probe modulation transfer only occurs at the very center of the focus where the intensity of the saturation beam is close to zero. Sub-diffraction-limited images can be obtained by raster-scanning the three collinearly aligned beams simultaneously across the sample. In the first experiment by Cheng's group[38] using the light path in **Fig. 3**b, the sub-diffraction imaging capability of about 225 nm on graphite nanoplatelets was demonstrated.

To further suppress the PSF and improve the resolution, the Wang group[40] used a frequency-doubling crystal to generate visible spectrum light (451 nm) for saturation excitation and demonstrated a resolution down to 36 nm. The substitution of saturation photons from near IR to the visible spectrum serves two functions. Firstly, the shorter wavelength naturally forms a smaller donut-shaped ring. Secondly, the visible spectrum can more effectively clear the electronic population in the valence band of graphene.

**Nonlinear differential TA microscopy**

It's also possible for TA techniques to achieve subwavelength resolution without the requirement of optical transition saturation. Differential TA microscopy can measure the difference between transient absorption signals induced by two pumps, as shown in **Fig. 3**. When the pump intensity is below the nonlinear threshold $p_0$ (**Fig. 3**c(I)), the differential TA is possible with the alternation of Gaussian and doughnut pumps, with an improvement in spatial resolution determined by the doughnut pump node dimensions. Super-resolution is further realized from the additional spatial frequencies (with respect to those defined by optical diffraction) that are introduced in the amplitude PSF when exploiting TA nonlinearity, for example by using two Gaussian pumps in alternation, one of power below the nonlinear threshold $p_0$ and one just above (**Fig. 3**c(II)), and thereby confining the nonlinear response to dimensions less than the pump beam diameter. Using CdSe nanobelts as samples, Liu et al. [42] demonstrated nonlinear differential TA microscopy with a resolution of 185 nm (λ/4.1NA) in the absence of TA saturation.

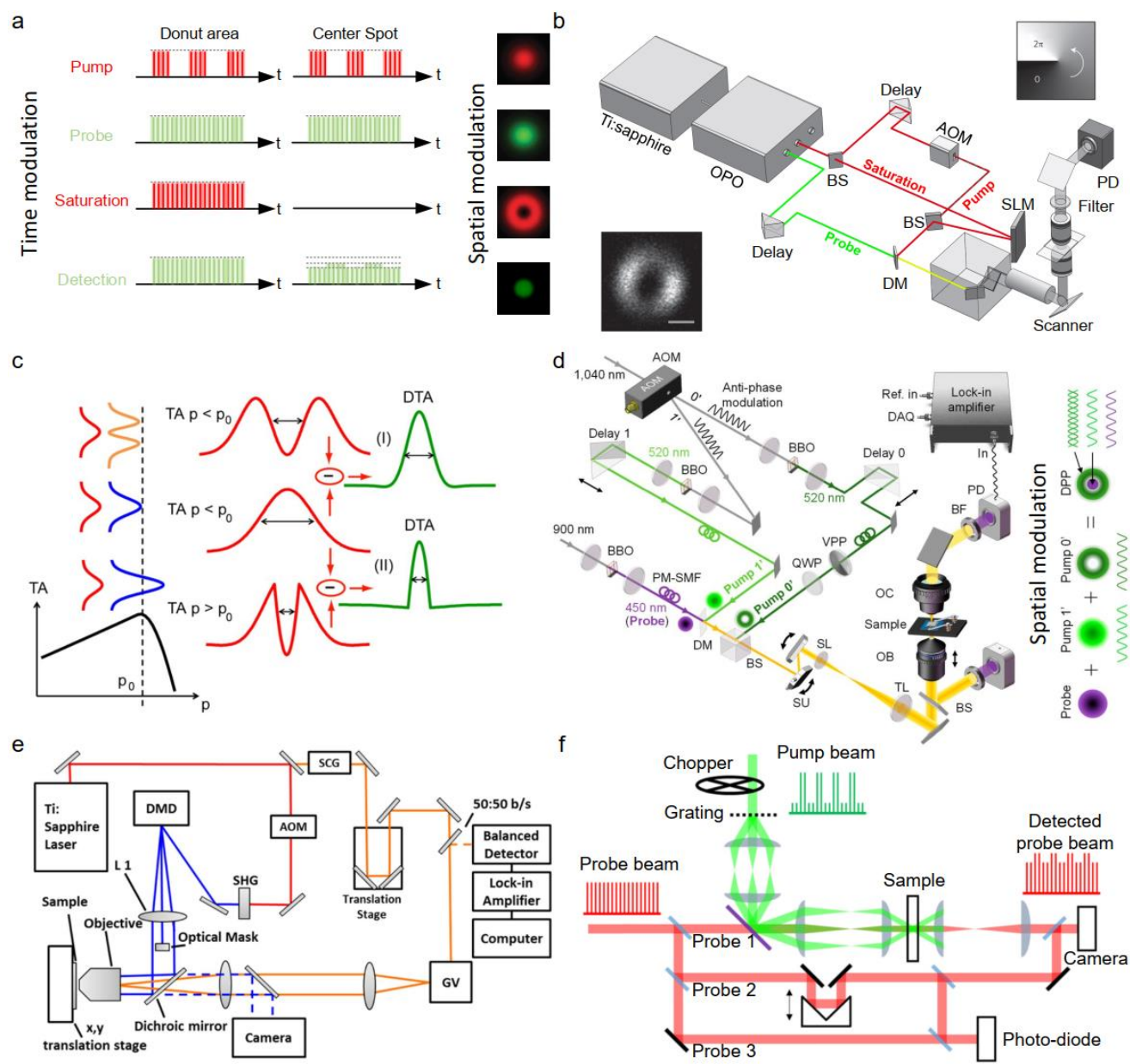

**Fig. 3 Implementation of various TA super-resolution microscopies. a** Principle of depletion-based TA microscopy. **b** Depletion-based transient absorption microscopy. Reproduced with permission[38].

Copyright 2013 Nature Publishing Group. **c** Nonlinear differential transient absorption microscopy. Reproduced with permission[42]. Copyright 2016 American Chemical Society. **d** Antiphase demodulation transient absorption microscopy. Reproduced with permission[41]. Copyright 2021 American Chemical Society. **e** Structured transient absorption microscopy. Reproduced with permission[43]. Copyright 2016 American Chemical Society. **f**. 3D structured pump-probe microscopy. Reproduced with permission[86]. Copyright 2017 Optical Society of America. OPO, optical parametric oscillator; AOM, acousto-optic modulator; BS, beam splitter; SLM, spatial light modulator; PD, photodiode; DM, dichroic mirror.

**Antiphase demodulation TA microscopy**

To achieve super-resolution imaging of interweaved copper wires, Yang et al.[41] impressed an additional pump laser in a donut shape with antiphase modulation and thus form a modulated subdiffraction-limit focus center in the laser focus, as shown in **Fig. 3**d. They have achieved a direct optical inspection of integrated circuits (ICs) with a lateral resolution down to 60 nm.

**Structured illumination TA microscopy**

The structured illumination TA microscopy employs a modulated pump excitation field to provide ultrafast spectroscopic measurements of sub-diffraction-limited sample areas. A diffraction-limited probe pulse is spatially overlapped with the pump field (**Fig. 3**e), and the sample is raster scanned to produce a pump-probe image at a well-defined delay time, $\Delta t$. The effective pump-probe excitation field is spatially modulated since the induced polarization of the sample depends on the product of the pump intensity and the probe field. The high spatial frequency information is encoded in the structured illumination images and can be recovered using a 2D coherent theoretical model[87]. A nearly two-fold reduction in 2D spatial resolution was achieved experimentally, from 223 nm to 114 nm by imaging the silicon nanowire[43]. The structured transient absorption microscopy could also be extended to 3D super-resolution by combining 3D coherent imaging theory with pump-probe microscopy[86](**Fig. 3**f).

The TA super-resolution microscopy could also be applied in reflection mode by photo-exciting temperature and/or carrier changes and probing of reflectance changes. The nonlinear components of reflectance with respect to photoexcitation allow the dramatic narrowing of the effective PSF. To extract the nonlinear components of reflection due to photo-modulation (at $\omega_m$), Tzang et al.[88, 89] demodulate the reflection intensity at the corresponding harmonic frequencies ($\omega_m$, $2\omega_m$, $3\omega_m$...) in a lock-in amplifier. The $n$th harmonics components of the reflectivity scale with the $n$th power of the excitation. Accordingly, the related effective $PSF_{pp}$ comprises the product of $PSF_{probe}$ and that of the pump laser to the power of the nonlinearity order $n$, $PSF_{pump}^n$. The width of the $PSF_{pp}$ scales down by $\sqrt{n}$ for Gaussian-focused beams. By scanning over the sample and measuring the nonlinear photo-modulated reflectivity, spatial resolution is enhanced beyond the diffraction limit. This methodology has been demonstrated on Si nanostructures and phase transition material such as vanadium oxide ($VO_2$)[88] with a spatial resolution down to 85 nm[89].

# 4. Raman scattering super-resolution chemical microscopy

Raman microscope is a versatile vibrational technique that is complementary to the IR microscope. Raman microscope detects the inelastically scattered light (Stokes) from vibrations which involve a change in the molecular polarizability. As water is a weak Raman scatterer, Raman microscopy is naturally suitable for living specimens. Coherent Raman scattering (CRS) microscopies, including

SRS[90] and CARS[91], are emerging techniques that improve the Raman imaging speed by enhancing the Raman signal through matching the beating frequency between pump and Stokes beam with the molecular vibration frequency. Achieving high-resolution Raman imaging with rich chemical information is very attractive in imaging live cells and even tissues. For CRS using near-infrared (NIR) light as excitation source (λ: 0.8-1.064 μm, NA: 1.49), the lateral spatial resolution can only reach ~300 nm[92]. Taking advantage of nonlinear optics, visible SRS microscopy shows improved spatial resolution with visible-wavelength pump and Stokes beams. By doubling the frequencies of the pump and Stokes beams to the wavelength around 450 nm, a spatial resolution of ~130 nm was achieved with a high-power oil immersion objective (NA = 1.49) [44, 45]. Despite these advances, the spatial resolving capability is still limited by Abbe's diffraction barrier. Below we broadly overview several super-resolution Raman imaging techniques using, structured illumination, signal blinking effect, depletion (or saturation) effect, high-order nonlinearity, and sample expansion.

**Structured illumination Raman**

Among the different super-resolution techniques, the structured illumination method is an ideal first choice for spatial resolution improvement in Raman microscopy as it does not impose special optical properties, such as switching, on the choice of samples. The principle of structured illumination is to project a fine illumination pattern on a sample to expand spatial frequencies resolvable by the optics without spoiling the analytical advantage of Raman microscopy.

In the structured-line illumination-Raman (SLI-Raman) [46] shown in **Fig. 4**a, a 532 nm CW laser is split by a phase grating to produce interference fringes along the slit illumination. Three fringe phases are varied by moving the position of the grating using a piezo scanner to reconstruct super-resolution in this line direction. With this system, Watanabe et al. achieved a 1.4-fold improvement in resolution along the slit illumination direction compared to the theoretical limit of a wide-field Raman microscope. They have demonstrated this improved resolution on polymer beads, graphene sheets, and fixed mouse brain slices.

The structure illumination Raman[47] could also be implemented in the wide-field scheme by combining wide-field SIM with a Raman band selecting module inserted between the objective lens and the tube lens(**Fig. 4**c). The Raman band selecting module is composed of a pair of tunable filters, which could be mechanically rotated to select the Raman spectrum captured by the camera. The high spatial resolution image can be reconstructed at every Raman band. The authors demonstrated the G-band imaging of the single-walled carbon nanotubes and hyperspectral imaging of graphene with SI-Raman with a spectral resolution of 50 cm$^{-1}$ and double spatial resolution improvement over the diffraction limit.

Compared with point-scanning Raman or line-scanning Raman(**Fig. 4**b), of which every pixel or column of pixels contains the full spectrum of the sample with a high spectral resolution, SI-Raman(**Fig. 4**d) is more time efficient but sacrifices the spectral resolution. Therefore, it's quite difficult to maintain both high acquisition speed and high spectral resolution at the same time. Besides, the spatial resolution of the conventional SIM method is limited to half of the diffraction limit because the illumination optics are also diffraction-limited, which restricts the finest period of the illumination pattern. Utilizing an objective lens with a higher NA combined with a shorter wavelength excitation source is one way of solving this. For example, an ultrahigh 1.7-NA objective and 266 nm ultraviolet light can bring the lateral resolution down to 40 nm. However, the increase of NA and magnification also bring disadvantages of smaller field-of-view, lower imaging depth, as well as high costs.

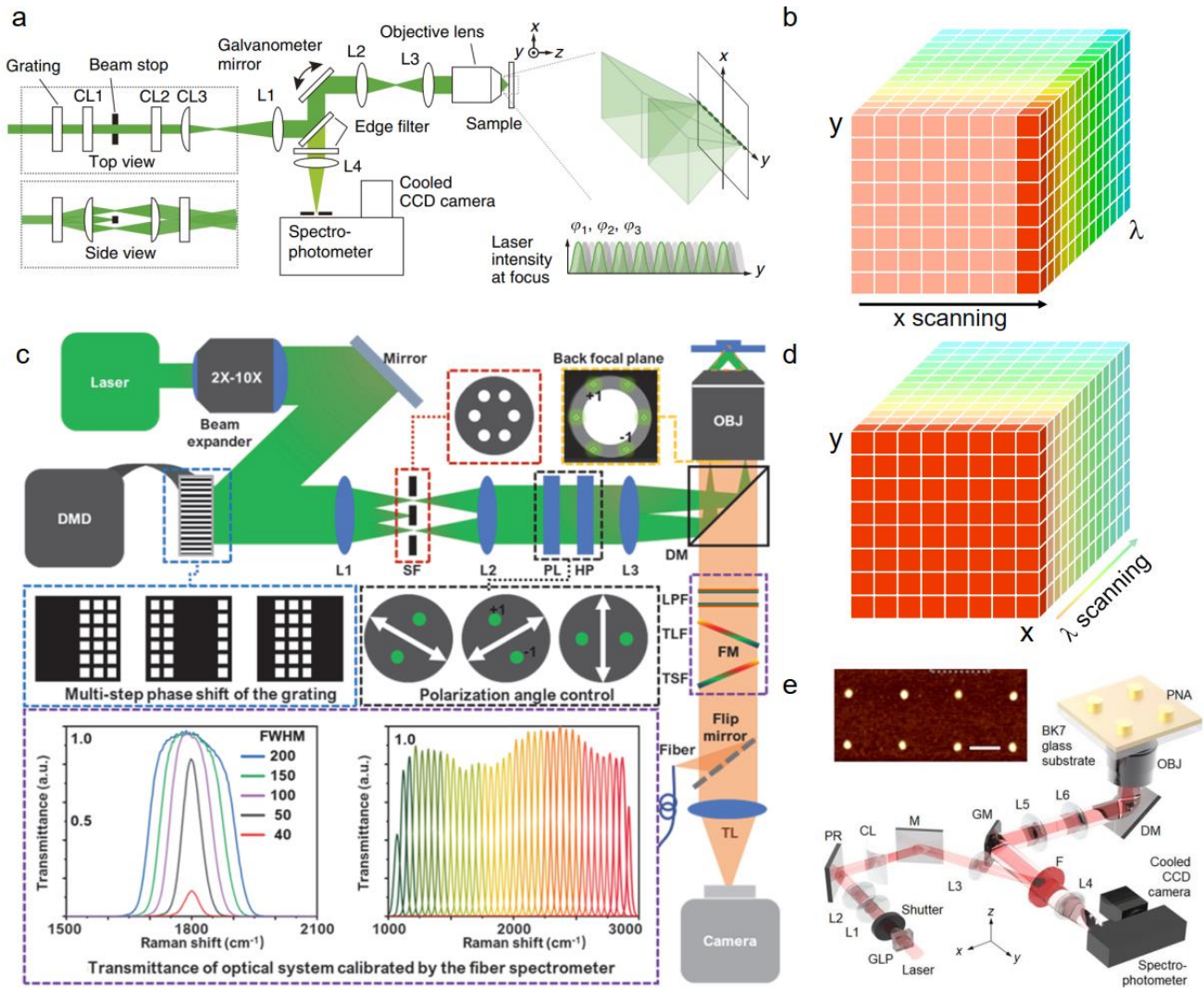

**Fig. 4 Optical implementation of various SI-Raman microscopies. a** Optical system of SLI-Raman microscopy. Reproduced with permission[46]. Copyright 2015 Nature Publishing Group. **b** 1D-SR hyperspectral imaging by scanning one spatial dimension. **c** Wide-field SI-Raman microscopy. Reproduced with permission[47]. Copyright 2021 American Chemical Society. **d** 2D-SR imaging by wavelength scanning. **e** SI-SERS. Reproduced with permission[48]. Copyright 2020 American Chemical Society.

One way of solving this problem is to detach the illumination path from the imaging path and replace the illumination module with photonic chips. The idea has been demonstrated in fluorescent microscopy, which is termed chip-based SIM (cSIM) [10, 11, 93]. The photonic chip is designed to generate evanescent patterned illumination with a much higher lateral wavevector than that by a conventional objective lens and thus can deliver a high resolution down to 65 nm[10] in fluorescent microscopy. The plasmonic nanostructures also support an enhanced near-field structured illumination and could be combined with Raman microscopy to improve both spatial resolution and signal sensitivity, termed SI-SERS[48, 49]. Lee et al. used gold nanopost arrays with a diameter of 100 nm and period of 500 nm as the substrate for Raman imaging, as indicated in **Fig. 4**e. The gold nanopost arrays can generate localized surface plasmons serving as the periodically structured light illumination. Common to the SIM method, the resolution enhancement arises from the subwavelength-period illumination, which provides a spatial-frequency-shift magnitude inversely proportional to the period. Therefore, the resolution can be enhanced by fabricating smaller period plasmonic arrays. Besides, the localized surface plasmons also have an enhanced field intensity, which bears a similar function as SERS. However, both the photonic chips and the plasmonic substrate have a limited illumination depth, which hinders the application in chemical imaging of biological specimens.

A computational scheme[57] was derived for reconstructing super-resolution CARS images combined with nonlinear coherent. The results demonstrate the method promises a benefit on CARS microscopy by adding the super-resolution capability to improve its 2D spatial resolution by a factor of approximately three.

**Nonlinear Raman**

Another way of further improving the resolution is to exploit non-linear properties of the material, such as the saturable absorption[94] and photoswitching[95]. Super-resolution SRS imaging in biological samples based on a saturated SRS technique has recently been demonstrated by combing with the virtual sinusoidal modulation (VSM) method[50]. The nonlinear signal can also be detected by point-scanning process with a result of squeezed PSF. The higher-order nonlinear process ($\chi(5)$, $\chi(7)$) was demonstrated to dominate over the cascaded lower-order nonlinear process under a tight focusing, which brings a much smaller excitation volume, giving rise to a significantly improved spatial resolution in the so-called HO-CARS microscopy[51]. The spatial resolution of ~190 nm has been realized by HO-CARS microscopy on unlabeled biological samples with 10 times decreased excitation power compared with saturated SRS technique[50, 96 97].

**Single-molecule localization microscopy (SMLM)-based Raman**

SMLM was proposed to overcome light's diffraction barrier using photoactivated or photo-switchable molecules to resolve the high density of molecules with super-resolution. This approach employs stochastic activation of fluorescence to switch on individual molecules and then images and bleaches them, temporally separating molecules that would otherwise be spatially indistinguishable. Merging all the single-molecule positions obtained by the photoactivation and imaging/bleaching cycles yields a final super-resolution image. The essential part of SMLM lies in the fluorophores that could be photoactivated or reversibly photo-switched by irradiation with light, such as photoactivatable or photoconvertible fluorescent proteins (PALM and FPALM) [4, 98]; pairs of organic fluorophores as activator and reporter (STORM) [5], or conventional fluorescent probes (dSTORM) [99, 100].

The application of the SMLM principle in label-free microscopy is quite restricted to the scarcity of blinking behavior. However, one exception is in SERS, which experimentally shows a rapid blinking behavior in the Raman spectral signal enhancement instead of signal emission (**Fig. 5**f). With this similar behavior of the fluorescent blinking in SMLM, one can use the time-dependent SERS signal for reconstructing chemically resolved, label-free super-resolution stochastic imaging (**Fig. 5**a). The possible mechanism of the blinking behavior in the SERS signal is assigned to be nanoscale local heating and thermal activation.

The localization precision of point-like objects in two dimensions can be described as [101]:

$$\sigma_x^2 = \frac{r_0^2 + q^2/12}{N} + \frac{8\pi r_0^4 b^2}{q^2 N^2}$$

where $r_0$ is the standard deviation of the point spread function, $N$ is the total number of photons collected, $q$ is the size of an image pixel, and $b$ is the background noise per pixel. For single blinking events resulting in ~$10^5$ visible photons, the position localization is expected to be on the order of a few nanometers.

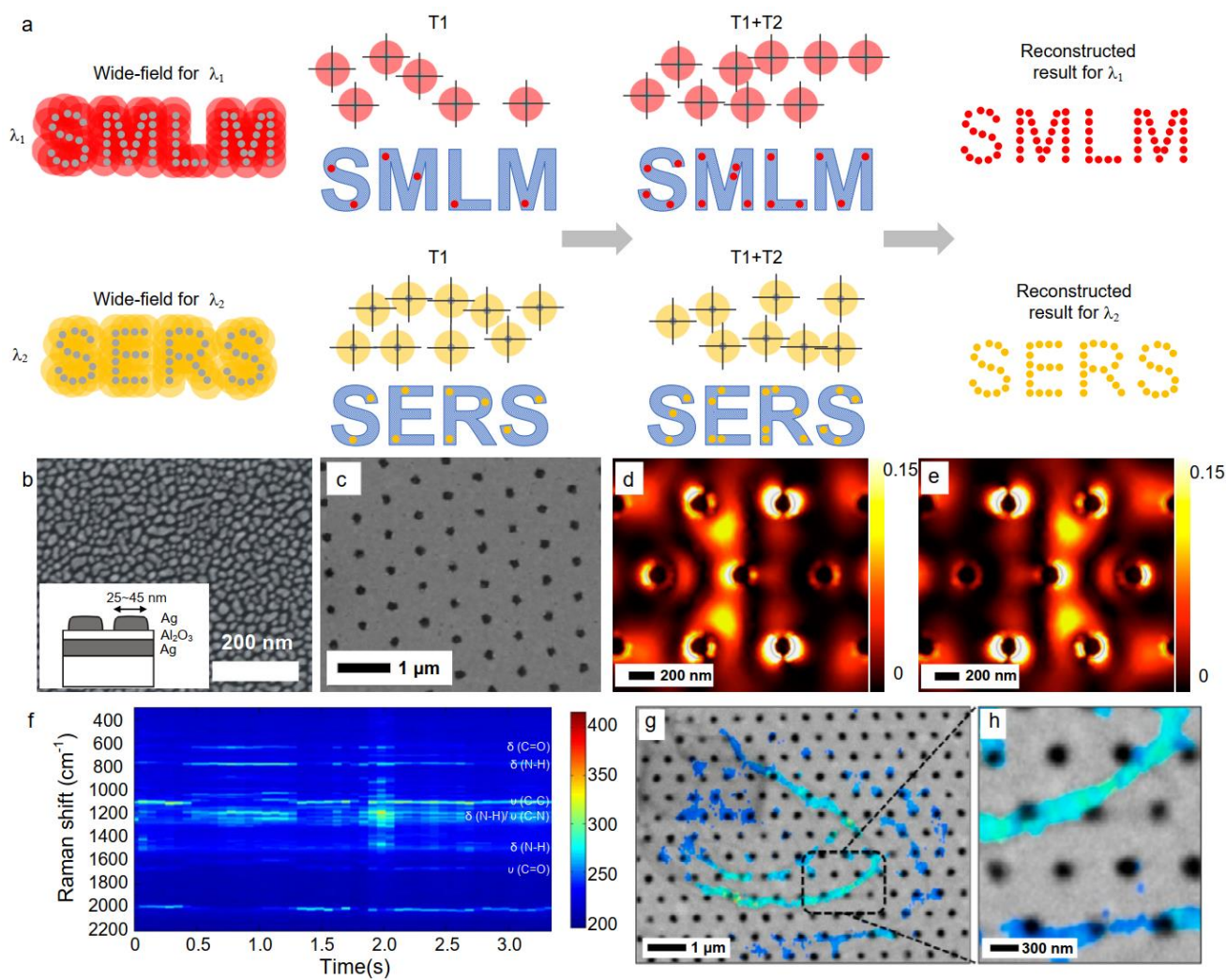

**Fig. 5 SMLM-based Raman. a** The principle of SMLM-based Raman. **b** Scanning electron micrograph of the self-organized metasurface for SERS substrate. Inset shows the schematic description of the substrate section. **c-e** Nanohole array chip for SERS substrate (c) and hotpot map generated by the substrate (d-e). Adjusting the phase of the six illuminating spots shifts the central hotspot to the left (d) to the right(e). (c-e) is reproduced with permission[53], Copyright 2014 American Chemical Society. **f** Typical SERS spectra recorded from a spot during wide-field SERS video imaging. (b, f) is reproduced with permission[52], Copyright 2013 Springer Nature. **g** Super-resolution SMLM-SERS imaging of collagen fibrils on a hexagonal silver nanohole array, overlayed with the same SEM image. **h** The enlarged image shown in the dashed box of (g). (g-h) is reproduced with permission[54], Copyright 2016 American Chemical Society.

In 2013, Ayas and Cinar et al.[52] reported the first SMLM-SERS by using metal-insulator-metal spontaneously organized metasurfaces for a substrate (**Fig. 5**b). The Ag film forms nano-islands with a mean particle diameter of 32 nm, which is designed to match the plasmon resonance wavelength close to the working laser wavelength of 532 nm. They exploited the SMLM-SERS to image self-assembled peptide networks and demonstrated Raman imaging with both high resolution (30 nm) and high sensitivity. The SERS substrate can also be single layer Ag (deposited on the sample)[52], a hexagonal array of nanoholes[53, 54] (**Fig. 5**c), or plasmonic nanopost arrays[48]. To alleviate the sampling imperfection caused by nonuniform static hotspot illumination, further improvements were obtained with dynamic illumination technique--by randomly altering the phase profile of the incident beam with a spatial light modulator (SLM) [53] or simple optical diffuser[54], with which the hotspots can be dynamically shifted (**Fig. 5**d-e). The SERS-STORM imaging method has been applied to collagen fibrils and has shown sufficiently accurate correlation with the results obtained from SEM (**Fig. 5**g-h).

SMLM-SERS can provide subdiffraction resolution but have difficulty providing quantitative information, and the plasmonic materials needed may affect reaction dynamics or cause sample degradation. Although the signal collection of SERS is in the far-field, another problem of the method is the plasmonics enhancement locates in the near-field (a few nanometers) of the substrate, which limits the application to only samples directly in contact with the plasmonic surface, such as cell membranes. Besides, the large number of raw images (a few thousand) required for reconstructing a high-resolution image also imposes limitations on the imaging speed.

**Depletion-based Raman**

As a fluorescence-based nanoscopy[102, 103], STED can achieve sub-50 nm resolution by shrinking the PSF of the imaging system using the nonlinear stimulated emission response of fluorophores. In the fluorescence-based STED, double beams of laser are used, one is a Gaussian beam for exciting the fluorophores, and the other is a doughnut-shaped 'STED' beam tuned in wavelength for silencing (depleting) the peripheral fluorophores in the diffraction-limited excitation regions via stimulated emission. The emission wavelength of the STED beam is red-shifted compared to that of the excitation laser and spontaneous emission fluorescence, thus only excited fluorophores located at the very center are selectively allowed to spontaneously emit fluorescence photons, the intensity distribution of which determines the PSF and spatial resolution of the system. As a result, by scanning the coaligned beams, a high-resolution image can be formed by stacking together fluorescent signals emitted from the very central region, of which the spatial resolution can be increased with higher saturation intensity.

For STED imaging, the critical technique is the way to switch off the signal light, which for a long time wasn't found in Raman imaging. In 2015, Silva et al. [55] proposed a similar way in STED to realize the depletion-based Raman based on femtosecond stimulated Raman spectroscopy (FSRS) [104], which is a nonlinear four-wave mixing technique utilizing vibrational coherences to generate stimulated Raman signals. The proposed method uses a broadband probe beam (830−1000 nm) and a picosecond pump beam (centered at 800 nm) to generate the Gaussian-shaped SRS signal along with a femtosecond doughnut-shaped depletion beam at the pump wavelength for turning off the outer edges of the Gaussian-shaped SRS signal. The detected Raman signal has a different wavelength from the probe light and could be filtered out and sent into a spectrograph for detection. The depletion of Raman signal from this method is nonlinear with increased depletion power, resulting in significant resolution enhancement of up to 47% when scanning the edge of a Raman-active diamond. The actual suppression mechanism was stated to be an "alternative four-wave mixing" pathway. The suppression of CARS signal can also be realized by using a three-beam double SRS scheme[105, 106], in which two different SRS processes by pump-Stokes and pump-depletion beam pairs are involved to competitively consume a limited number of common pump photons.

**Sample-expansion Raman**

While most optical super-resolution microscopies focused on manipulating illumination laser or fluorophores to bypass the diffraction limitation, expansion microscopy (ExM) physically expands the specimens by embedding them in a swellable polymer network[107]. Anchored to the gel, fluorescent probes[107], proteins[108, 109], RNA[110], DNA[111], and lipids[112, 113] can be isotropically separated in space to greater distances, and therefore can be effortlessly revolved even by conventional diffraction-limited microscopes. For example, a resolution of 25-70 nm can be achieved with a conventional confocal microscope[114], according to the expansion factors of 4-10×.

Since the rise of sample expansion, ExM has been constantly iterating with many variations (MAP[115], proExM[108], iExM[116], click-ExM[117], U-ExM[118], pan-ExM[119], Cryo-ExM[120, 121]). Meanwhile, the combination of expansion microscopy with other super-resolution microscopies (SIM[122], STED[123, 124], STORM[125], FED[126]) and SRS[56, 127] have been demonstrated. The combination of ExM with SRS, which is termed vibrational imaging of swelled tissues and analysis (VISTA), achieves a decent three-dimensional SRS imaging with a resolution down to 78 nm through optimal retention of endogenous proteins, isotropic sample expansion, and deprivation of scattering lipids[56](**Fig. 6**b). The VISTA imaging mainly targets the methyl group (i.e., $CH_3$) vibrations from endogenous proteins at 2940 $cm^{-1}$ for visualizing protein-rich structures (**Fig. 6**a). Another example[127] uses methacrolein and heat denaturation, instead of 6-((acryloyl)amine) hexanoic acid (AcX) and strong protease digestion, to achieve ~ 100% protein retention rate during the sample expansion. With this method, up to 10× expansion and sub-50 nm resolution, SRS nanoscopy can be realized with multiple Raman and fluorescent probes tagged.

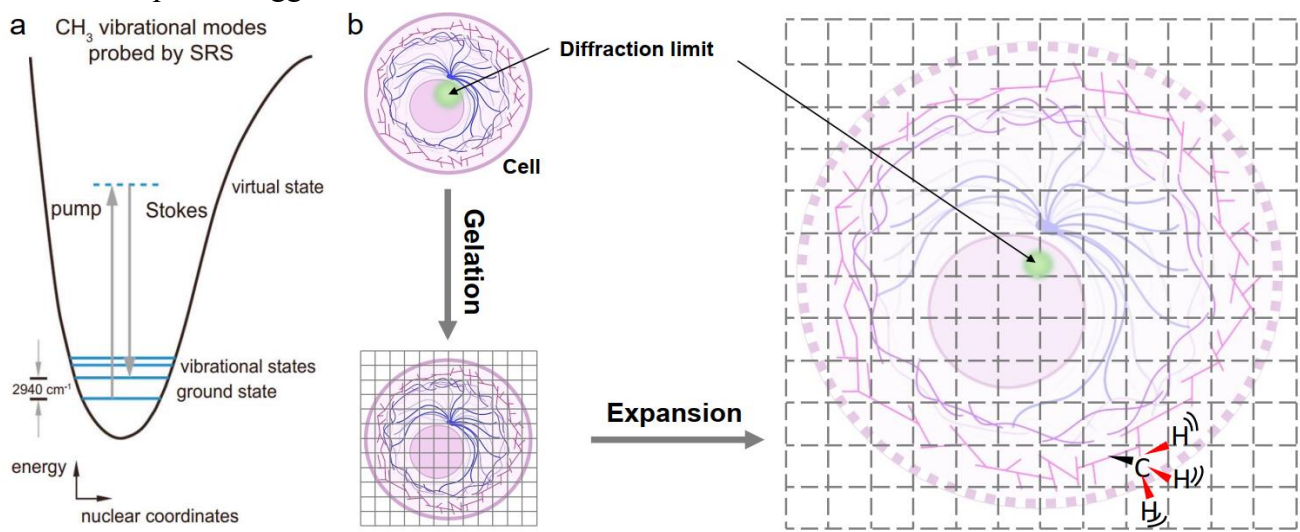

**Fig. 6 The principle of sample-expansion Raman microscope. a** Energy scheme for SRS probing of vibrational motion (taking $CH_3$ channel at 2940 $cm^{-1}$ as an example). Reproduced with permission[56]. Copyright 2021 Nature Publishing Group. **b** General sample treatment process of expansion microscopy, including fixation, gelation, and expansion.

## 5. Applications of far-field super-resolution chemical microscopy

Since the absorption-based and Raman scattering-based super-resolution chemical microscope can provide subdiffraction-limit chemical information encoded in intrinsic chemical bond vibrations, they can be easily applied—either individually or in combination—to broad applications ranging from biology, pharmaceutics, histopathology, material science, environmental contamination detection, cultural heritage conservation, and industrial application such as ICs inspection, as indicated in **Fig. 7** and **Fig. 8**.

## 5.1 Biological and medical applications

SRCM has found broad applications in label-free imaging of intrinsic molecules, such as DNA, RNA, protein, lipid, and glucose, in biomedical samples, from viruses and cells to tissues, as highlighted below.

**Microorganism detection and classification**

Virus is a kind of microorganism consisting of a nucleic acid molecule (DNA or RNA) and a protective coat (protein) in sub-micrometer size. The detection of the virus, such as SARS-CoV-2[128], represents both a worrisome public health problem and an increasingly common source of therapeutics and vaccines. Taking advantage of the high sensitivity and high specificity, vibrational spectroscopies like Raman spectroscopy have been applied for discriminating current or past infection by SARS-CoV-2 in molecular diagnostics[129]. As another vibrational spectroscopy, interferometric scattering-MIP can achieve chemical imaging of single viruses with the size of 80 nm, thanks to the high sensitivity and high resolution. The 1650 $cm^{-1}$ for the amide I band and 1,550 $cm^{-1}$ for the amide II band of viral proteins were used for mapping poxvirus and vesicular stomatitis virus (VSV). The intensity ratio between the amide I band and amide II band was studied to discriminate different virus types[69], as shown in **Fig. 7**a. This study experimentally demonstrates the label-free imaging of single viruses with fingerprint spectral information.

Besides, the rapid and accurate classification and discrimination of bacteria, especially at the single-cell level, is an important task[130] in microorganism. SRCM is an attractive technique for meeting this requirement benefitting from the advantages of high resolution, high speed, high reliability, and low cost. The MIP imaging of various bacteria, such as S. aureus (**Fig. 7**b) and E. coli has been obtained at both fingerprint region and high wave number region[66, 67, 131-133]. The different spectral features across the amide I and II bands in varying bacteria were also demonstrated, potentially applicable in high-throughput single bacteria characterization and classification.

**Cell biology**

The cellular compositions and metabolism of living cells[134, 135] were mapped with SRCM. Lipid metabolism is closely associated with many metabolic diseases, such as atherosclerosis, cancer, obesity, and diabetes. Lipid droplets stored PC-3 prostate cancer cells[31] (**Fig. 7**c), bladder cancer T24 cells[35] (**Fig. 7**d), S. cerevisiae yeast cells(**Fig. 7**f), and SKOV3 human ovarian cancer cells[32] were mapped around the frequency of 1,750-$cm^{-1}$ corresponding to the vibrational peak of the C=O bond by MIP. The fatty acid uptake of A549 lung cancer cells was imaged at high-wavenumber using MIP[136]. While for Raman imaging, bonds of C=C, $CH_2$ in lipid (2,845 $cm^{-1}$) are usually used as the imaging target for higher signals (**Fig. 7**e). Also, SRS can differentiate C=C in lipid (1,650 $cm^{-1}$) and cholesterol (1,669 $cm^{-1}$).

For proteins, another key player in cellular activity, the frequency of protein amide I band at 1,656 $cm^{-1}$ was typically chosen to map the protein contents[32, 35] (**Fig. 7**d, f).

**Digital histopathology and pathologies**

SRCMs have also drawn great interest in digital histopathology[137, 138] based on the inherent signature of the tissue samples. With SLI-Raman, the 1D super-resolution spontaneous Raman map of the mouse brain slice has been obtained (**Fig. 7**g). The 1,682 cm$^{-1}$ (red) and 2,848 cm$^{-1}$ (green) correspond to amide-I and CH$_2$ stretching vibrational modes in the brain slice. MIP-based histopathology has the potential to provide real hematoxylin and eosin (H&E) results with similar visible wavelength spatial resolutions. Schnell et al. demonstrated the whole breast tissue microarray slide with MIP imaging at discrete IR wavelengths and developed the classification algorithm to differentiate cell subtypes[137], as shown in **Fig. 7**h.

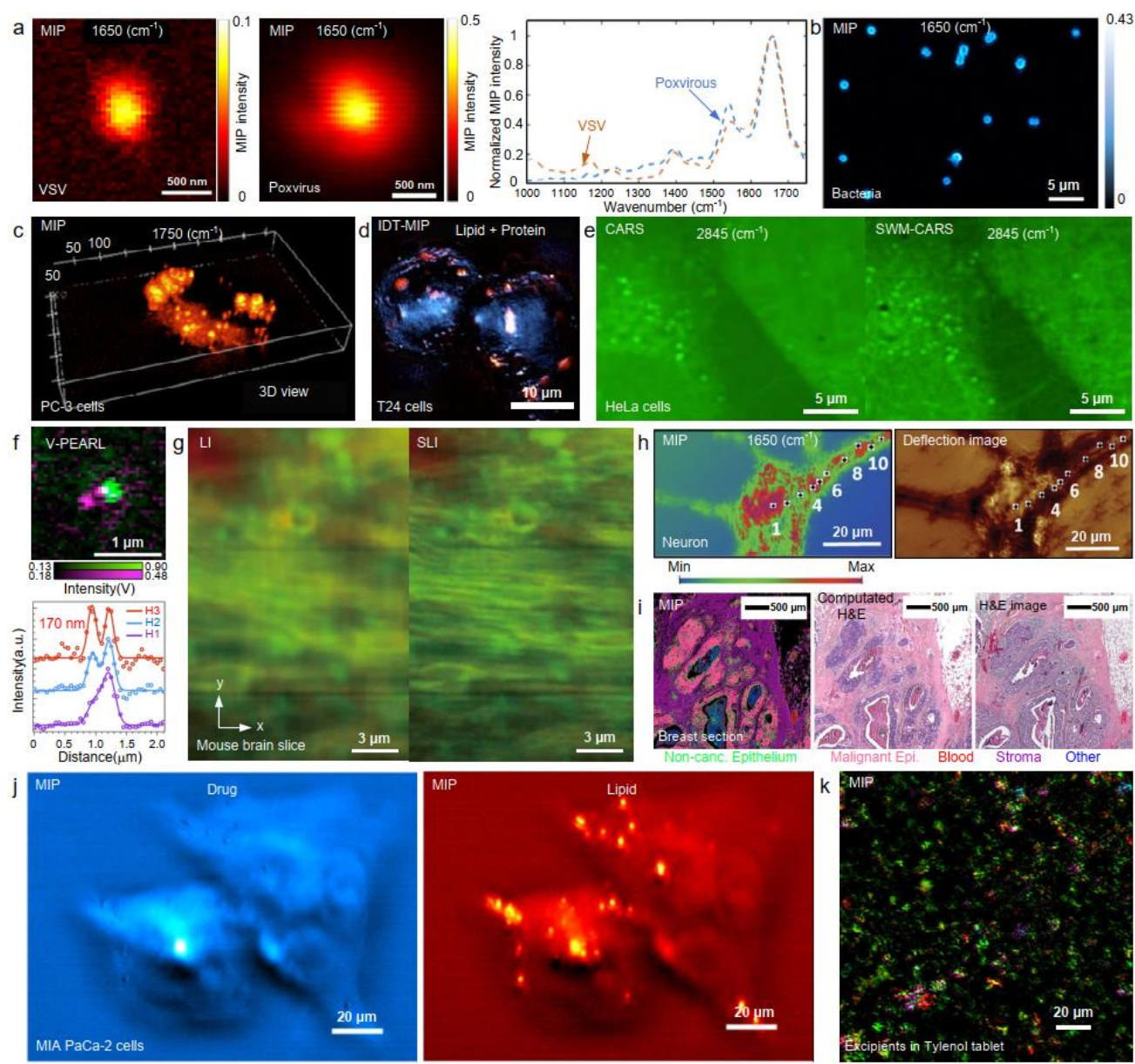

**Fig. 7 Applications of super-resolution chemical microscopy in biological and medical studies. a** Interferometric scattering MIP image of single virus (left: VSV, right: poxvirus) with the IR laser tuned to 1,650 cm$^{-1}$ (amide I band). Scale bars: 500 nm. The amide I to amide II intensity ratio showed a difference for various virus types. Reproduced with permission[69]. Copyright 2021 American Chemical Society. **b** MIP images of single bacteria at Amide I (1650 cm$^{-1}$). Reproduced with permission[133]. Copyright 2020 American Chemical Society. **c** 3D MIP imaging of PC-3 cells at the 1,750 cm$^{-1}$ C=O band. Reproduced with permission[31]. Copyright 2016, the Authors, published by AAAS. **d** 3D MIP chemical imaging of fixed bladder cancer T24 cells under mitosis (red: lipid, green: protein).

Reproduced with permission[35]. Copyright 2022 Nature Publishing Group. **e** CARS and SWM-CARS images of HeLa cells at 2,845 cm$^{-1}$(CH$_2$ stretching of lipids). Scale bars: 5 μm. Reproduced with permission[51]. Copyright 2019 Nature Publishing Group. **f** Overlapped V-PEARL image of live S. cerevisiae yeast cells. red: 1,650 cm$^{-1}$ amide I band, green: 1,750 cm$^{-1}$ lipid C=O band. Reproduced with permission [37]. Copyright 2023 Nature Publishing Group. **g** Raman images of a mouse brain slice using line illumination (LI) and structured-line illumination (SLI). The intensity distribution of Raman peaks at 1,682 cm$^{-1}$ (red) and 2,848 cm$^{-1}$ (green) can be assigned to amide-I and CH$_2$ stretching vibrational modes, respectively. Scale bar, 3 μm. Reproduced with permission[46]. Copyright 2015 Nature Publishing Group. **h** MIP image of the neuron acquired at 1650 cm$^{-1}$ showing the distribution of proteins (left), deflection image of a primary neuron(right). Scale bars: 20 μm. Reproduced with permission[139]. Copyright 2020 The Authors. Published by WILEY-VCH. **i** MIP histopathology of an unstained breast surgical resection, combining automated recognition and traditional pathology. (left) Classification of surgical resection from MIP data and (middle) its derived computational H&E. (right) H&E image of the adjacent section. Scale bars: 500 μm. Reproduced with permission[137]. Copyright 2020, the Authors, published by National Academy of Sciences. **j** MIP imaging of JZL184-treated MIA PaCa-2 cells for drug (left) and lipid (eight) content. Scale bars, 20 μm. Reproduced with permission[31]. Copyright 2016, the Authors, published by AAAS. **k** Epi-MIP image of the distribution of API (green), corn starch (red), PVP (cyan), and SSG (magenta) in a Tylenol tablet. Scale bars: 50 μm. Reproduced with permission[71]. Copyright 2017 American Chemical Society.

SRCM shows the potential for studying brain pathologies, such as amyloid plaques in Alzheimer's disease. MIP was demonstrated to characterize amyloid structures present inside neurons at sub-micrometer resolution. Evidence of localized β-sheet elevations and the polymorphic nature of β-sheet structures at the subcellular level in AD transgenic neurons were provided[139, 140] by mapping protein distribution inside a primary neuron at 1650 cm$^{-1}$, as shown in **Fig. 7**i. By combining MIP with synchrotron-based X-ray fluorescence (S-XRF), iron clusters were found to co-localize with elevated amyloid β-sheet structures and oxidized lipids, which motivates the application of high-resolution multimodal micro-spectroscopic approaches in the study of pathologies[140].

**Drug mapping**

Visualizing pharmaceutical compounds in the regulation of cellular activity in living systems is important to understand their mechanism of action. SRCM can be applied for mapping drug molecules distribution in living cells[31] (**Fig. 7**j) and testing the distribution of active pharmaceutical ingredients in tablets[71, 141](**Fig. 7**k).

**5.2 Material science, environmental study, cultural heritage conservation, and chip inspection**

SRCM has also found applications in material science, environmental contamination detection, cultural heritage conservation, and ICs inspection, where labeling methods are difficult to apply.

**Material science**

SRCM has widely been applied in the characterization of submicrometer materials such as graphene[38-41, 142](**Fig. 8**a), nanotubes[41](**Fig. 8**c), nanowires[43, 54] (**Fig. 8**d), and nanobelts[42]. For typical artifacts like nanowrinkles in monolayer graphene (**Fig. 8**b), the size is only tens of nanometers, which could be easily detected by TA super-resolution microscopy[40]. As another important optoelectronic material, halide perovskite has promising performance characteristics for low-cost optoelectronic applications[143]. The photovoltaic performance of halide perovskite devices strongly depends on the composition distribution, which could be chemically mapped by MIP. With the help of MIP, spatially resolved local cation-related compositional inhomogeneities in mixed cation perovskite films were imaged[144]. In another research, cation migration under different bias voltages in lead halide perovskites devices was mapped using MIP[145] (**Fig. 8**e). These studies suggest a materials design strategy for suppressing cation instabilities in hybrid perovskites.

**Environmental contamination detection**

Micro(nano)plastics (MNPs) released from polymer-based products are ubiquitous in the environment and bring environmental and human health risks. SRCM has been demonstrated to be a decent tool for studying MNPs because of the sub-micrometer resolution, high sensitivity, and chemical imaging ability. With the help of MIP, Kniazev et al. [146] used the MIP to characterize MNPs released from nylon tea bags that had been steeped in water at different temperatures (**Fig. 8**g). They also chemically identify MNPs in sieved road dust. In another study, Su et al. examined the MNPs released from Silicone-rubber baby teats after steam disinfection by using MIP, which indicated that > 0.66 million elastomer-derived micro-sized plastics could be ingested by a baby by the age of one year[147] (**Fig. 8**f).

**Cultural heritage conservation**

SRCM has also been applied in mapping cultural heritage such as paintings[148, 149] (**Fig. 8**i) and glass-metal objects[150] (**Fig. 8**h). For paintings, the metal soaps are detrimental and will degrade the appearance and integrity. The root cause of metal soap formation was hindered by the limited spatial resolution of FTIR. With the help of a commercialized MIP, high signal-to-noise ratio and spatial resolution distribution of phase-separated species were mapped and analyzed[148]. Both crystalline zinc carboxylate phases (1530-1558 $cm^{-1}$, sharp peaks) and disordered Zn-soap phases (1550-1660 $cm^{-1}$, broad peaks) were found to coexist within the 0.1 $\mu m^3$ volume probed (**Fig. 8**i). This result proves that SRCM is a reliable tool for promoting the development of cultural heritage conservation practices.

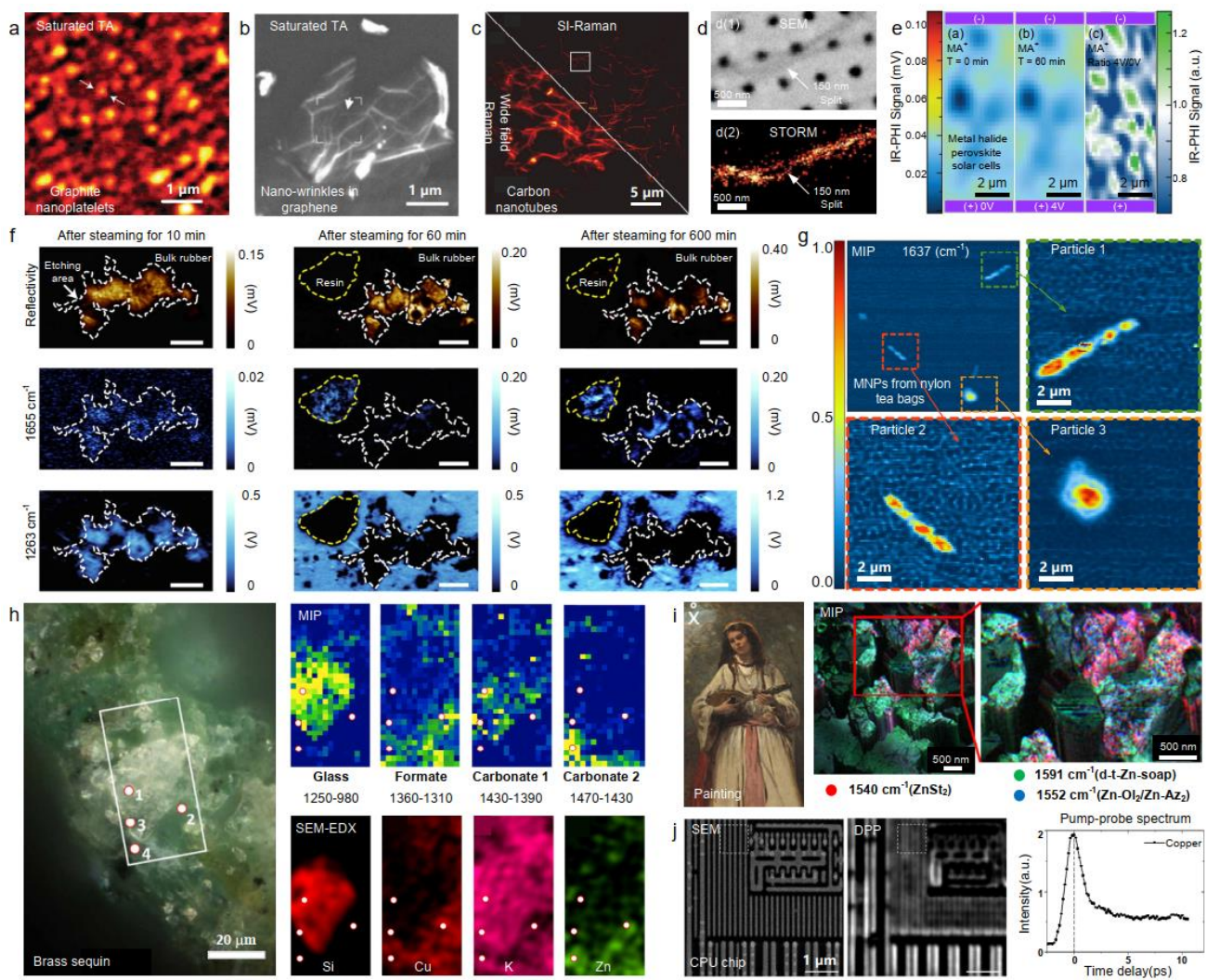

**Fig. 8 Applications of super-resolution chemical microscopy in material science, environmental study, cultural heritage conservation, and chip inspection. a** Saturated transient absorption microscopy imaging of graphite nanoplatelets. Reproduced with permission[38]. Copyright 2013 Nature Publishing Group. **b** Super-resolution imaging of nano-wrinkles in graphene. Reproduced with permission[40]. Copyright 2020 Optical Society of America. **c** SI-Raman imaging of single-walled carbon nanotubes. (c1) Wide-field Raman image. (c2) SI-Raman reconstructed image. Reproduced with permission[47]. Copyright 2021, the Authors. Published by American Chemical Society. **d** The SERS-STORM imaging technique reproduces the fine features of the strand, seen to traverse between the nanoholes and, in other regions to have small splits. Reproduced with permission[54]. Copyright 2016 American Chemical Society. **e** Analogous MIP signal maps of the MA$^+$(1450 cm$^{-1}$) NH$_3$ symmetric bend before and after 60 minutes of applied bias (4V, |E|=0.1 V/μm) in metal halide perovskite solar cells. Reproduced with permission[145]. Copyright 2020 American Chemical Society. **f** Visible-laser and MIP imaging of C=O (1655 cm$^{-1}$) and Si−CH$_3$ (1263 cm$^{−1}$) of one location on the teat subsample surface 10, 60 and 600 min after steaming. Scale bars, 100 μm. Reproduced with permission[147]. Copyright 2021 Nature Publishing Group. **g** MIP imaging of MNPs from nylon tea bags steeped at 25 °C. All images were analyzed at 1637 cm$^{−1}$. Reproduced with permission[146]. Copyright 2021 American Chemical Society. **h** High-resolution MIP imaging of degradation products on the surface of the degraded brass sequin, which matches well with the SEM-EDX elemental mapping.

Reproduced with permission[150]. Copyright 2022, the Authors, published by AAAS. **i** Reconstructed qualitative color-coded map of MIP absorption displaying ZnSt$_2$ (1540 cm$^{-1}$, red), tetrahedral Zn carboxylate (1591 cm$^{-1}$, green), and ZnOl$_2$/ZnAz$_2$ (1552 cm$^{-1}$, blue) of a French 19th-century painting. Reproduced with permission[148]. Copyright 2022 American Chemical Society. **j** DPP imaging of layer of copper interconnections in a CPU chip. Reproduced with permission[41]. Copyright 2021 American Chemical Society.

**Chip inspection**

SRCM can also be applied in the inspection of ICs in central processing unit (CPU) chips[41]. Using an antiphase demodulation pump-probe (DPP) microscope, direct super-resolution imaging with a lateral resolution of 60 nm was demonstrated on imaging the multilayered copper interconnects in CPU (**Fig. 8**j). Compared with conventional ICs inspection methods, such as scanning electron microscopy (SEM), transmission electron microscopy (TEM), atomic force microscopy (AFM) [151], ptychographic X-ray computed tomography (PXCT) [152], and ptychographic X-ray laminography (PyXL) [153], SRCM method opens the possibility of providing easy, fast, and large-scale ICs inspections.

## 6. Summary and perspective

SRCMs are capable of revealing the chemical landscape of cells (i.e., the composition and distribution of molecules as well as the dynamical interactions) or materials at the submicrometer level without tagging or genetically encoding fluorescent labels. Up to now, the best resolution performance was achieved to be 10 nm by SMLM-SERS, among various far-field SRCMs. In parallel, STED-TA achieved a sub-50 nm resolution by inducing a visible saturation laser[40].

However, a trade-off exists between resolution and other imaging performances, such as imaging depth, acquisition speed, field of view, and spectrum channels. Such a trade-off arises from the limited information throughput. For example, SMLM-SERS and SI-SERS could achieve a high resolution but sacrifice the imaging depth for using evanescent illumination. In addition, for these methods, more acquisition time is required when pursuing a better resolution. Continuing developments in spectroscopy and microscopy will render SRCM a more powerful chemical-imaging platform in the foreseeable future.

The detection sensitivity undermines the SNR and imaging speed especially when the resolution is down to the nanometer scale. In chemical microscopy, the sensitivity is fundamentally limited by the shot noise induced by randomness in the times that photons are detected. For samples with weak or without characteristic vibrational features, one possible way of improving the sensitivity is to apply small tags[70, 154, 155] for SRCM imaging. Another possible way of solving this problem is quantum photon correlations[156]. The experiment has been performed by combining coherent Raman scattering microscope with quantum correlation technique, which allows imaging of molecular bonds within a cell with a 35 percent improved SNR compared with conventional microscopy[157].

To further push the resolution, the key lies in the integration of chemical microscopy with newly developed super-resolution techniques or algorithms. Raman scattering-based chemical microscopies have been physically studied and combined with various super-resolution methods such as SIM, SMLM, and STED. However, the application of these SRCMs is either limited by high illumination intensity or a long acquisition time. Recently reported deconvolution algorithm-based super-resolution SRS microscopy indicated the possibility of further reducing the resolution down to sub-60 nm by using the newly developed algorithm[158]. While the IR absorption-based SRCMs have achieved much success and found many applications from biomedical study to material science, the best resolution achieved is still limited to the probe light (around 300 nm[36, 66]). We expect MIP could be integrated with SIM to break the visible diffraction limit.

Moreover, scatter-based and absorption-based SRCMs can be combined with each other or other chemical imaging techniques to provide complementary information on the samples[131]. For example, secondary ion mass spectrometry (SIMS)[159] and X-ray fluorescence[160, 161] can be combined with SRCM to correlate elemental, isotopic, and molecular information in two or three dimensions[162]. As a highly efficient way to excite molecular vibration, SRCM can also be coupled with other physical processes and observables, such as optical coherence tomography[163, 164], optical diffraction tomography[165], photo-induced force microscopy[166], atomic force microscopy[151, 167], and electron microscopy[168], which could offer benefits in terms of penetration, resolution or imaging contrast.

Lastly, the large-scale application requires high-throughput multispectral imaging, which will benefit from the advanced spectral reconstruction methods, combing with computational algorithms such as deep learning method[169]. Quantitative analysis, such as multi-variate curve resolution, previously used for the de-composition of hyperspectral stimulated Raman scattering data, can be applied to extract both the spectral profile and the concentration map of chemical content.

Looking forward, we envision that advanced SRCM imaging and analysis will be a major force in future biological, medical, chemical, and material discovery.

**Acknowledgments:** This work was supported in part by the National Natural Science Foundation of China (No. T2293751, 62020106002, 92250304, 31901059) to Q.Y. and Y. H.

**Author contributions**
All the authors contributed to the writing of the manuscript.

**Conflicts of Interest**: The authors declare no conflict of interest.

microalgae measured by synchrotron radiation X-ray fluorescence and absorption microspectroscopy. *Environmental Science & Technology* **50**, 8827-8839 (2016).
162. Decelle, J. *et al.* Subcellular chemical imaging: new avenues in cell biology. *Trends in Cell Biology* **30**, 173-188 (2020).
163. Israelsen, N. M. *et al.* Real-time high-resolution mid-infrared optical coherence tomography. *Light: Science & Applications* **8**, 11 (2019).
164. Siddiqui, M. *et al.* High-speed optical coherence tomography by circular interferometric ranging. *Nature Photonics* **12**, 111-116 (2018).
165. Dong, D. S. *et al.* Super-resolution fluorescence-assisted diffraction computational tomography reveals the three-dimensional landscape of the cellular organelle interactome. *Light: Science & Applications* **9**, 11 (2020).
166. Davies-Jones, J. A. & Davies, P. R. Photo induced force microscopy: chemical spectroscopy beyond the diffraction limit. *Materials Chemistry Frontiers* **6**, 1552-1573 (2022).
167. Ando, T., Uchihashi, T. & Fukuma, T. High-speed atomic force microscopy for nano-visualization of dynamic biomolecular processes. *Progress in Surface Science* **83**, 337-437 (2008).
168. Goodhew, P. J., Humphreys, J. & Beanland, R. Electron Microscopy and Analysis. 3rd edn. (New York: Taylor & Francis, 2001).
169. Huang, L. Q. *et al.* Spectral imaging with deep learning. *Light: Science & Applications* **11**, 61 (2022).